\documentclass[
aps
prd,
preprint,
superscriptaddress,
amsfonts,
amssymb,
amsmath,
preprintnumbers,
floatfix,
nofootinbib,
unsortedaddress,
showpacs,
tightenlines,
floats,
a4paper,
eqsecnum,
]{revtex4}
\def\dhc{\bar{\delta}_\m{hc}}
\def\Ki{K_\m{i}(r)}

\def\al{\alpha}
\def\be{\begin{equation}}
\def\ee{\end{equation}}
\def\ep{\epsilon}

\def\fr{\frac}
\def\de{\delta}
\def\ga{\gamma}
\def\de{\delta}
\usepackage{color}
\usepackage{comment}

\def\al{\alpha}
\usepackage[dvipdfmx]{graphicx}
\def\be{\begin{equation}}
\def\ee{\end{equation}}
\def\ep{\epsilon}

\def\fr{\frac}
\def\de{\delta}

\def\ga{\gamma}
\def\Ga{\Gamma}
\def\de{\delta}

\def\al{\alpha}
\def\si{\sigma}

\def\m{\mathrm}

\def\ri{r_\m{i}}

\makeindex

\begin{document}

\title{The double formation of primordial black holes}

\author{Tomohiro Nakama}

\affiliation{Department of Physics,
Graduate School of Science,\\ The University of Tokyo, Bunkyo-ku,
Tokyo 113-0033, Japan
}

\affiliation{Research Center for the Early Universe (RESCEU),\\
Graduate School of Science, The University of Tokyo, \\ Bunkyo-ku,
Tokyo 113-0033, Japan
}

\date{\today}

\preprint{RESCEU-39/14}

\date{\today}

\small

\begin{abstract}
Primordial black holes (PBHs) are a useful tool in cosmology to probe
primordial inhomogeneities on small scales 
that reenter the Hubble radius during the radiation dominated epoch. 
In this paper, a phenomenon we call the double formation of PBHs, described below, is explored. 
Suppose there exists a highly perturbed region which will collapse to form a PBH after the horizon crossing of this region, 
and farther that this region is superposed on much 
larger region, which also collapses upon reentry. 
One then expects the collapse of the central smaller region at the time of the crossing of this 
region, followed by another collapse of the larger perturbation at the time of its respective crossing. The smaller PBH, formed earlier, should be swallowed in the second collapse leading to a single 
larger PBH as the final state. 
This paper reports the first direct numerical confirmation of such double PBH formation. 
Related to this, we also discuss the effects of high-frequency modes on the formation of PBHs.

\end{abstract}
\maketitle
\section{introduction}
Any large amplitude perturbation of order unity 
 can collapse to a primordial black hole (PBH) in the radiation-dominated universe 
after the horizon crossing
 \cite{Zel'dovich-1974,Hawking:1971ei}. 
The mass of PBHs is of the order of the horizon mass at the horizon crossing and 
therefore can span a wide range between $\sim 10^{-5}$g and $\sim 10^5M_{\odot}$\footnote
{
Recently, constraints on PBHs as the seeds of 
supermassive and intermediate mass black holes are discussed \cite{Kohri:2014lza}. 
}. 

PBHs with different masses 
can lead to different observational signals, 
leading to a vast amount of literature in this field. 
So far there has been no conclusive evidence of the existence of PBHs, 
and observational upper bounds on the abundance on various mass scales have been obtained, 
which are updated and summarized in
\cite{Carr:2009jm} together with new constraints imposed by Hawking radiation \cite{Hawking:1974rv}. 
In the following, we mention several of these. 

PBHs smaller than $\sim 10^{15}$g would have evaporated by now
through Hawking radiation, emitting high-energy particles, 
which can change the history of 
Big-Bang Nucleosynthesis \cite{Zel'dovich:1977sta,Novikov:1979pol,
1978AZh....55..231V,1978PAZh....4..344V,Miyama:1978mp,Kohri:1999ex}. 
In addition, these high-energy particles could be observed as 
 the gamma-ray
background \cite{Page:1976wx,MacGibbon:1987my,MacGibbon:1991vc} or 
galactic and extragalactic antiprotons \cite{1976ApJ...206....8C}.

Larger PBHs would still exist today and are
constrained by lensing effects \cite{Paczynski:1985jf}. 
Also, PBHs can be constrained by the stochastic gravitaional wave background for the following reason. 
When PBHs are formed in large numbers, 
the typical amplitude of primordial inhomogeneities is faily large $\sim {\cal O}(10^{-1})$ \cite{Bugaev:2008gw,Josan:2009qn} 
(note that the amplitude is $\sim {\cal O}(10^{-5})$ on largerst observable scales $\sim \mathrm{Gpc}$, relevant 
to CMB or large-scale structure). 
In this case, large amplitude tensor perturbations are generated due to the second-order effects of 
scalar perturbations (induced gravitational waves), 
potentially leading to observational imprints in the stochastic gravitational wave background
\cite{Saito:2008jc,Saito:2009jt}.  

PBHs are important partly because they may explain dark matter. 
Up until recently, there has been a "window" of mass scales of PBH dark matter, $10^{20}$g-$10^{26}$g \cite{Carr:2009jm}, 
namely, PBHs in this mass range have been a viable candidate of dark matter. 
However, more recently, progress has been made in the observational constraint of PBHs in this window. 
Several authors have claimed that the possibility of PBHs being all the dark matter has been excluded 
based on micro-lensing \cite{Griest:2013aaa} and PBH 
capture by neutron stars \cite{Capela:2013yf,Pani:2014rca} 
and stars \cite{Capela:2014ita}. 
Therefore, the window for PBH dark matter has been narrowed substantially. 
However, this is not the whole story. 
Several authors have claimed that Hawking radiation stops when the mass of PBHs reaches the Planck mass, 
with so called "Planck mass relics" left, 
and these relics can also provide a viable candidate of dark matter \cite{MacGibbon:1987my} (see also \cite{Carr:2009jm} and references therein). 
It would be fair to say that the relative importance of Planck mass relics being dark matter 
has been increased since the aforementioned window has been diminishing, and so 
further exploration of this possibility is much awaited. 

Observational upper bounds on PBHs 
provide valuable information on inflationary models
\cite{1989LNP...332..369P,Sato:1980yn,Guth:1980zm,Starobinsky-1980}, predicting
generation of super-horizon curvature perturbations
\cite{Mukhanov:1982nu,Guth:1982ec,Hawking:1982cz,Starobinsky:1982ee}.
So far, primordial perturbations on largest observable scales have been precisely probed by 
CMB \cite{Hinshaw:2012aka,Ade:2013uln} and
large-scale structure \cite{Bird:2010mp}. 
It is equally important to probe 
primordial inhomogeneities on significantly smaller scales in order to help pinpoint the correct 
inflationary model. Indeed, there exist a number of 
inflationary models that predict the enhancement of the power spectrum at small scales
\cite{PhysRevD.54.6040,
PhysRevD.42.3329,PhysRevD.50.7173,yokoyama-1997-673,
PhysRevD.58.083510,Jun'ichi1998133,kawasaki-1999-59,
PTPS.136.338,1475-7516-2008-06-024,PhysRevD.59.103505,
PhysRevD.63.123503,
PhysRevD.64.021301,Kawasaki:2007zz,Kawaguchi:2007fz,Kohri:2007qn,Alabidi:2009bk,
Alabidi:2012ex,Kawasaki:2012kn,Kawasaki:2012wr}. 
PBHs provide one of the important proves of primordial inhomogeneities on small scales 
(others include CMB distortions \cite{Chluba:2012we}, ultracompact minihalos \cite{Bringmann:2011ut}, 
acoustic reheating \cite{Nakama:2014vla}). 

In order to use PBHs as a reliable tool in cosmology, precise knowledge of their formation condition 
is required. 
A considerlation of the balance between gravity and
pressure gradients, which hamper contraction, yielded a simple analytic criterion of PBH formation $1/3\lesssim \dhc\label{classical}$ \cite{Carr:1974nx,Carr:1975qj}, 
where $\dhc$ is the energy density perturbation averaged over the 
overdense region evaluated at the time of horizon crossing in the uniform Hubble slice. 
This criterion has long been used in predicting PBH abundance
(but has recently been analytically refined in \cite{Harada:2013epa}).
In this simple picture 
the dependence on the 
profile or shape of perturbed regions has not been taken into account.

Recent numerical analyses, however, have shown that the condition for PBH formation does
 depend on the profile of perturbation
 \cite{Shibata:1999zs,Polnarev:2006aa} (see also \cite{1979STIN...8010983N,PhysRevD.59.124013} for earlier work). 
In these papers, some functions were introduced to model primordial perturbed regions 
and obtained conditions for PBH formation. 
However, their functions include at most two parameters and therefore
the types of initial perturbation profiles investigated were limited. 
In reality, various kinds of perturbations must have been generated during inflation so 
in our previous paper \cite{1475-7516-2014-01-037} (hereafter NHPY) we considered a considerably wider class of perturbations 
by introducing a function including as many as five 
parameters. 
For this extended class of shapes, we have found 
that the condition for PBH formation is generically expressed by two quantities 
characterizing profiles of perturbations; one is presented as an integral of curvature
over initial configurations and the other is presented in terms of the position of the boundary and
the edge of the core, which may measure the effects of pressure gradients. 

It turns out that the function introduced in NHPY enables us to investigate a phenomenon we call the double formation of PBHs. 
Suppose there exists a highly perturbed region which will collapse to form a PBH after horizon crossing, 
and also that this region is superposed on a much 
larger region, which also collapses as it enters the horizon later. 
Then, 
the collapse of the central smaller region at the time of the crossing 
should be followed by another collapse of the larger perturbation at the time of the crossing of this 
larger perturbation. The smaller PBH, formed earlier, is involved in the second collapse leading to a 
larger PBH as the final state. It is expected that the first collapse is not significantly affected by the presence of the larger perturbation 
since it is still outside the horizon at the time of the crossing of the smaller perturbation 
\footnote{
Relate to this issue, in \cite{Young:2014ana}, 
it is argued that one should focus on the density contrast on comoving slices to correctly calculate PBH abundance, 
considering the existence of super-horizon modes of the curvature perturbation.
}. 
This paper is aimed at reporting a first direct numerical confirmation of this phenomenon of double PBH formation 
\footnote{
This phenomenon is an analogue of situations
where a dark matter halo, formed at some time, becomes a part of a larger halo later, in the process of large-scale structure formation. 
These situations are 
taken into account in Press-Schechter formalism \cite{Press:1973iz}.
This issue has been discussed in the literature in the context of PBHs as well, for example in \cite{Carr:1975qj}. 
But it may be interesting to note the difference between the halo and PBH case. 
In the case of dark matter halos, a given halo is \textit{destined} to be involved in a larger halo 
and this process takes place continuously; a halo forms at some time and in the next instant this halo 
becomes a part of a slightly larger halo. This is because perturbations of the dark matter always grow and 
the formation of a halo is determined solely by whether the amplitude of the density perturbation, smoothed over each scale, exceeds the threshold value $\sim 1.68$ (for the spherical case), 
irrespective of the timing (in this case the notion of the horizon crossing does not play any role since the halo formation takes place well inside the horizon). 
For the case of PBHs, the double-(or potentially multiple-)formation does not always take place, and when it happens it happens basically discretely 
(the next PBH formation, involving another smaller PBH or PBHs inside, takes place after some finite time interval). 
This is because whether a perturbation collapses to form a PBH has already been determined by the time of the horizon crossing, and if it does not collapse, it disperses completely. 
}. 

In the double formation of PBHs, from the smaller PBH perspective, one is simply swallowed by the larger PBH. 
But from the point of view of the larger PBH, the presence of the smaller-scale perturbation leading to the smaller PBH 
corresponds to the existence of a high-frequency mode (hereafter a HF mode), whose wavelength is much shorter than the perturbed region under consideration. 
In numerical simulations of the formation of PBHs, 
the presence of HF modes 
has not been taken into account \cite{Shibata:1999zs,Polnarev:2006aa,1979STIN...8010983N,PhysRevD.59.124013,1475-7516-2014-01-037}. 
In reality, HF modes should also exist 
unless PBHs result from a spike in the primordial power spectrum with extremely small width, 
and thus affect the formation of PBHs to some extent so 
investigating this issue is also important to fully understand the dynamics of the PBH formation. 
This has been numerically investigated for the first time and we find that 
HF-modes facilitate the formation of PBHs.

The rest of the paper is organized as follows. 
In \S I\hspace{-.1em}I, we briefly discuss our methods and 
\S I\hspace{-.1em}I\hspace{-.1em}I is dedicated to a 
discussion of the double formation of PBHs. 
In \S I\hspace{-.1em}V, 
the effects of high-frequency modes are discussed and  
\S V is devoted to the conclusion.

\section{Method}
We now review the methods employed. 
More details can be found in NHPY. 

The metric we employ is \cite{Misner:1964je}
\begin{equation}
ds^2=-a^2dt^2+b^2dr^2+R^2(d\theta^2+\sin^2\theta d\phi^2),\label{1}
\end{equation}
where $R$, $a$ and $b$ are functions of 
$r$ and the time coordinate $t$. We consider a perfect fluid with energy density $\rho(r,t)$ and pressure 
$P(r,t)$ 
with a constant equation-of-state parameter $\ga$ such that $P(r,t)=\ga \rho(r,t)$.
We express the proper time derivative of $R$ as
\begin{equation}
U\equiv \frac{\dot{R}}{a},\label{2}
\end{equation}
where dots denote derivatives with respect to $t$.

We define the mass, sometimes referred to as the Misner-Sharp mass in the literature, within the shell of circumferential radius $R$ by
\be
M(r,t)=4\pi\int^{R(r,t)}_0\rho(r,t)R^2dR.\label{defofM}
\ee
We consider the evolution of a perturbed region embedded in a 
flat Friedmann-Lemaitre-Robertson-Walker (FLRW) Universe with metric
\be
ds^2=-dt^2+S^2(t)(dr^2+r^2d\theta^2+r^2\sin^2\theta d\phi),
\ee
which is a particular case of (\ref{1}). The scale factor in this background evolves as
\begin{equation}
S(t)=\left(\fr{t}{t_\mathrm{i}}\right)^{\alpha} ,\quad\alpha\equiv\frac{2}{3(1+\gamma)},
\label{S0anddefofalpha}
\end{equation}
where $t_\mathrm{i}$ is some reference time.

We denote the background solution with a subscript 0.
In terms of the metric variables defined in (\ref{1}), we find
\be
a_0=1,\: b_0=S(t),\:  R_0=rS(t).
\ee
The background Hubble parameter is
\begin{equation}
H_0(t)=\fr{\dot{R_0}}{a_0R_0}=\frac{\dot{S}}{S}=\frac{\alpha}{t},
\end{equation}
and the energy density is calculated from the Friedmann equation,
\begin{equation}
\rho_0(t)=\frac{3\alpha^2}{8\pi Gt^2}.
\end{equation}
The energy density perturbation is defined as
\be
\de(t,r)\equiv \fr{\rho(t,r)-\rho_0(t)}{\rho_0(t)}\label{defofdelta}.
\ee
The curvature profile $K(t,r)$ is defined by writing $b$ as
\be
b(t,r)=\frac{R'(t,r)}{\sqrt{1-K(t,r)r^2}}\label{16}.
\ee
Note that this quantity $K(t,r)$ vanishes outside the perturbed region so that the solution asymptotically approaches the background FLRW
 solution at spatial infinity.

We denote the comoving radius of a perturbed region by $r_\mathrm{i}$, the 
precise definition of which will be given later (see eq. (\ref{defri})), 
 and define a dimensionless parameter
$\ep$ in terms of the squared ratio of the Hubble radius  $H_0^{-1}$ to the physical
length scale of the configuration,
\be
\ep \equiv \left(\frac{H_0^{-1}}{S(t)r_\mathrm{i}}\right)^2
=(\dot{S}r_\mathrm{i})^{-2}
=\frac{t_\mathrm{i}^{2\al}t^{\beta}}{\al^2 r_\mathrm{i}^2},\quad \beta\equiv 2(1-\al).\label{defofep}
\ee
When we set the initial conditions for PBH
formation, the size of the perturbed 
region is much larger than the
Hubble horizon. 
This means $\ep\ll 1$  
at the beginning, so it can serve as an
expansion parameter to construct an analytic solution of 
the system of Einstein equations to describe the spatial dependence of
 all the above variables 
at the initial moment when we set the initial conditions. 
In this paper, the second order solution, obtained in \cite{1475-7516-2012-09-027}, is used to provide initial conditions 
for the numerical computations.

We define the initial curvature profile as
\be
K(0,r)\equiv K_\mathrm{i}(r),
\ee
where $K_\mathrm{i}(r)$ is an arbitrary function of $r$ 
which vanishes outside the perturbed region.
Note that,
from (\ref{16}), $K_\mathrm{i}(r)$ has to satisfy the condition
\be
K_\mathrm{i}(r)<\fr{1}{r^2}.\label{condforKi}
\ee
We normalize the radial Lagrangian coordinate $r$ in such a way 
that $K_\mathrm{i}(0)=1$.

In order to represent the comoving length scale of the 
perturbed region,
we use the co-moving radius,
$r_\mathrm{i}$, of the overdense region.
We can calculate $r_{\mathrm i}$ by 
solving the following equation for the energy density perturbation defined by (\ref{defofdelta}):
\be
\delta(t,r_{\mathrm i})=0.\label{defri}
\ee
Since the initial condition is taken at the super-horizon regime, when
$\epsilon$ is extremely small, the following lowest-order solution \cite{Polnarev:2006aa} 
\be
\de(t,r)=\fr{2\ri^2}{9r^2}(r^3\Ki)'\ep(t)\label{firstorder}
\ee
suffices to calculate $r_\mathrm{i}$, which is obtained by solving 
\be
3K_\mathrm{i}(r_\mathrm{i})+r_{\mathrm i}K'_\mathrm{i}(r_\mathrm{i})=0. 
\label{rieq}
\ee
Note that the physical length scale in the
asymptotic Friedmann region
is obtained by multiplying by the scale factor $S(t)$, the 
normalization of which
we have not specified.  We can therefore  set up initial 
conditions for the PBH formation with arbitrary mass scales by adjusting
the normalization of $S(t)$ which appears in the expansion parameter.

The following equations were used in
\cite{MayWhite} to analyze the gravitational collapse of spherically symmetric masses:
\be
\dot{U}=-a\left(4\pi R^2\fr{\Ga}{w}P'+\fr{MG}{R^2}
+4\pi GPR \right)\label{u_t},
\ee
\be
\dot{R}=aU,\label{R_t}
\ee
\be
\fr{(\nu R^2)^{\cdot}}{\nu R^2}=-a\fr{U'}{R'}\label{rhoRsq},
\ee
\be
\dot{E}=-P\left(\fr{1}{\nu}\right)^{\cdot}\label{E_t},
\ee
\be
\fr{(aw)'}{aw}=\fr{E'+P(1/\nu)'}{w},\label{aw}
\ee
\be
M=4\pi\int^r_0\rho R^2R' dr, \label{mdef}
\ee
\be
\Ga=4\pi \nu R^2R',\label{Gamma}
\ee
\be
P=\ga \rho,\label{eos}
\ee
\be
w=E+P/\nu,\label{w}
\ee
where $E\equiv \rho/\nu$ and
\be
\nu\equiv\fr{1}{4\pi bR^2}\label{rhob}.
\ee
The constraint equation reads
\be
\left(\fr{R'}{b}\right)^2=\Ga^2=1+U^2-\fr{2M}{R}.\label{constraint}
\ee
Boundary conditions are imposed such that
$U=R=M=0$ and $\Ga=1$ at the center, 
and $a=1, \rho=\rho_0$ on the outer boundary
so that the numerical solution is smoothly connected to the 
FLRW solution. 

In this slicing, the computation stops after the horizon is formed due to the appearance of a singularity, 
so the eventual mass of the PBH can not be determined. 
The determination of the mass without facing a singularity through 
the technique of null slicing \cite{1966ApJ...143..452H,0264-9381-6-2-012,1995ApJ...443..717B,PhysRevD.59.124013,0264-9381-22-7-013} was also discussed 
in NHPY. 
In this slicing, space-time is sliced along the null geodesics of hypothetical photons 
emitted from the center and reaching a distant observer. 
In other words, the space-time is sliced with hyper-surfaces, 
defined by a constant null coordinate $u$, the so-called observer time defined shortly. 
By this construction of the null slicing, only information outside the horizon is calculated, 
without looking into what happens inside the apparent horizon. 
Initial conditions are given on some hypersurface defined 
by constant $u$ 
and are obtained using the cosmic time slicing by 
calculating the null geodesic of a hypothetical photon 
which reaches a distant observer after being emitted from the center at some moment in time,
while at the same time recording the physical quantities on this null geodesic 
\cite{1995ApJ...443..717B}. 
In this slicing, the information can be obtained without facing a singularity 
until a sufficiently later time 
when the eventual mass of a PBH can be determined.

Let us define the time variable $u$ by first 
noting
\be
adt=bdr\label{null}
\ee
along an outgoing photon. 
Then $u$ is defined by
\be
fdu=adt-bdr,\label{defu}
\ee
where $f$ is the lapse function necessary to make $du$ 
a perfect differential. 
From this definition, (\ref{null}) holds along 
the hyper-surfaces each defined by constant $u$, 
meaning that these surfaces correspond to the null geodesics of outgoing photons. 
Using $u$ as the time variable 
then means that the space-time is sliced with the null slices. 
A boundary condition on the lapse function is imposed 
by setting $a(u,r=\infty)=f(u,r=\infty)=1$, hence
$u=t$ at the surface defined by $r=\infty$. 
The physical meaning of this boundary condition is that $u$ is chosen to coincide with the proper time measured by 
a distant observer residing at spatial infinity in the background FLRW universe. 
For this reason, the null slicing is also sometimes referred to as observer time slicing in the literature. 

The Einstein equations in null slicing were obtained in 
\cite{1966ApJ...143..452H}, later used to simulate gravitational collapse followed by 
the formation of a black hole
\cite{0264-9381-6-2-012,1995ApJ...443..717B}, 
and to simulate the PBH formation as well
\cite{PhysRevD.59.124013,0264-9381-22-7-013}. 
We used numerical techniques similar to those of 
\cite{1995ApJ...443..717B,0264-9381-22-7-013}. 
The fundamental equations are as follows:
\be
U=\fr{1}{f}R_u \label{Rdot2},
\ee
\be
\fr{1}{f}M_u=-4\pi R^2PU \label{mwrtu},
\ee
\be
E_u=-P\left(\fr{1}{\nu}\right)_u \label{Ewrtu},
\ee
\be
b=\fr{1}{4\pi \nu R^2}\label{defrho},
\ee
\be
\fr{1}{f}U_u
=-\fr{3}{2}\left(\fr{4\pi \Ga R^2}{w}P'+\fr{M+4\pi R^3P}{R^2}\right)
-\fr{1}{2}\left(4\pi \nu R^2U'+\fr{2U\Ga}{R}\right)\label{Uwrtu},
\ee
\be
\fr{1}{f}\left(\fr{1}{\nu}\right)_u
=\fr{1}{\nu\Ga}
\left(
\fr{2U\Ga}{R}
+4\pi\nu R^2U'
-\fr{1}{f}U_u
\right)\label{rhowrtu},
\ee
\be
\fr{1}{b}\left(\fr{\Gamma+U}{f}\right)'=-4\pi R\fr{\rho+P}{f}\label{fderi},
\ee
where the subscript $u$ denotes differentiation with respect to $u$. 
Boundary conditions are the same as those in the cosmic time slicing.

\section{The double formation of PBHs}
In NHPY, the following function was introduced to parameterize various types of initial curvature profiles:
\be
K_\mathrm{i}(r)=A\left[1+B\left(\frac{r}{\sigma_1}\right)^{2n}\right]
\exp\left[-\left(\fr{r}{\sigma_1}\right)^{2n}\right]
+(1-A)\exp\left[-\left(\fr{r}{\sigma_2}\right)^{2}\right].\label{newfunction}
\ee

Before discussing the double PBH formation, let us first consider a simple case with $(A,B,\si_1,n)=(1,0,1.45,1)$ to demonstrate the 
results of a numerical computation for the case of a single PBH formation. 
\begin{figure}[h]
\begin{center}
\includegraphics[width=15cm,keepaspectratio,clip]{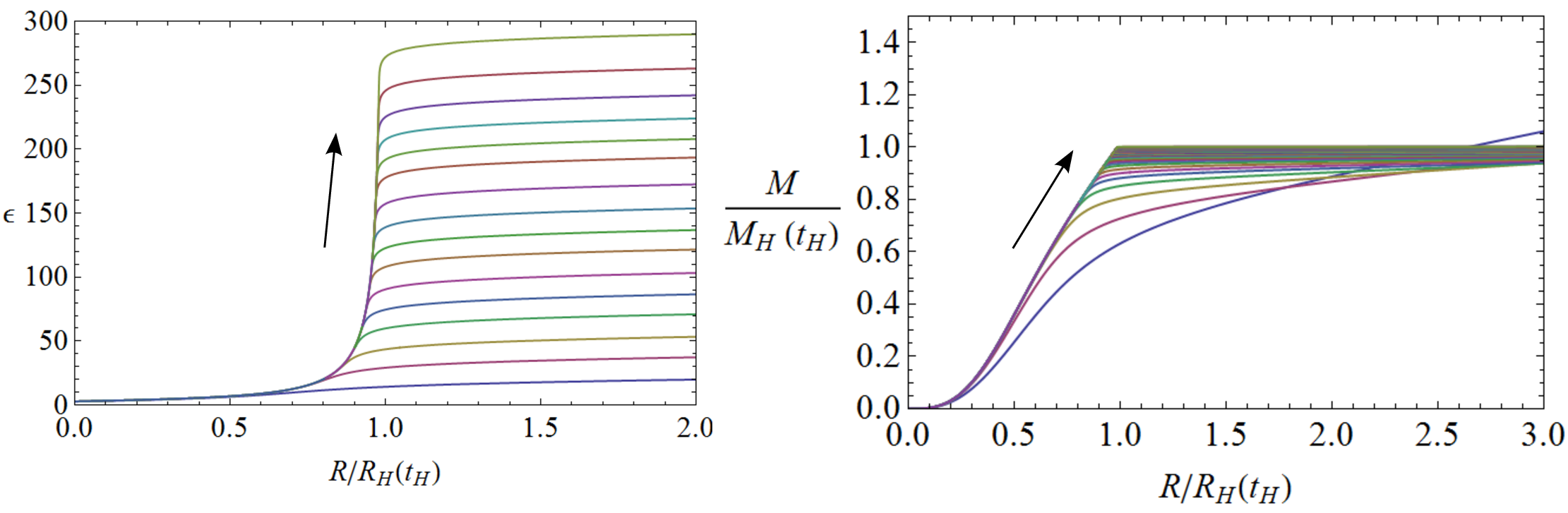}
\end{center}
\caption{
The results when $(A,B,\si_1,n)=(1,0,1.45,1)$. 
The left panel shows the null geodesics with the horizontal axis representing 
the circumferential radius $R$ normalized by the Hubble radius $R_{\rm{H}}(t_{\rm{H}})$
and vertical axis $\ep$, which can be regarded as a time variable. 
The right panel shows the time evolution of the mass profile, normalized by the 
horizon mass at the time of the horizon crossing $M_{\rm{H}}(t_{\rm{H}})$. 
Arrows show the direction of time evolution. 
}
\label{single}
\end{figure}
The hypersurfaces of $u=$const., corresponding to null geodesics, are shown in the left panel of 
Fig.\ref{single}. Observe that the intervals between null geodesics 
are tiny in the central region, meaning that time here is effectively frozen. 
Therefore, the formation of a singularity can be avoided in this slicing as expected. 
The upper lines in this figure correspond to the null geodesics 
of the hypothetical photons which are emitted from the center at later times and 
feel the effects of stronger gravity, 
so that they need more time to 
reach a distant observer. 
In this figure 
there is an envelope curve of the null geodesics, which 
approximately shows the location of the apparent horizon. 
In this way the time evolution is computed only outside the apparent horizon, 
so the eventual mass of a PBH can be determined without facing a singularity. 
From the same figure, the apparent horizon radius can be confirmed to asymptote to a constant value after its formation. 
This means that the black hole mass asymptotes to a constant value because 
$R=2M$ on the apparent horizon, 
and this behavior of the mass can be confirmed by the converging curves of the mass profile in the 
right panel of Fig.\ref{single}. 

The flatness of the mass profile in later times can be understood by noting 
that the energy density in a region away from the center decreases due to the expansion of the universe 
and also due to the existence of an underdense region surrounding the central overdense region so that the spacetime approaches the spatially flat FLRW universe. 
In this example, the eventual mass of the PBH is $\sim M_{\rm{H}}(t_{\rm{H}})$. 

\begin{figure}[h]
\begin{center}
\includegraphics[width=15cm,keepaspectratio,clip]{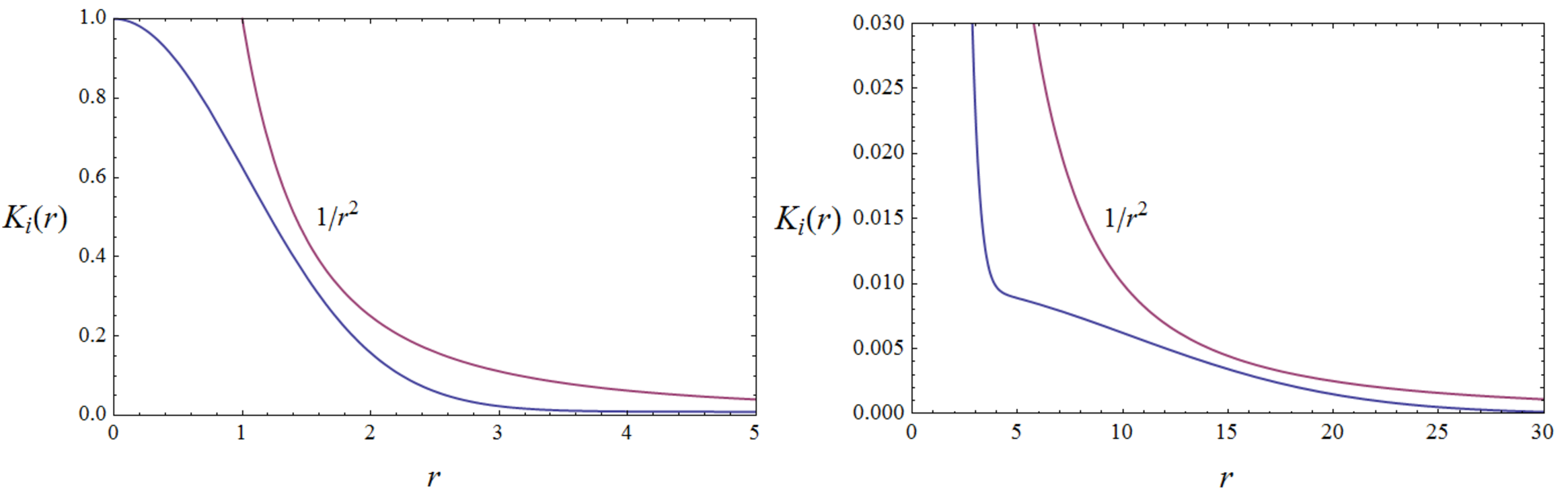}
\end{center}
\caption{
The initial curvature profile eq.(\ref{newfunction}) with $(A,B,\si_1,\si_2,n)=(0.99,0,1.45,10\si_1,1)$. 
The left panel shows the central perturbed region represented by the first term of eq.(\ref{newfunction}), 
which is superposed on the perturbed region shown in the right panel and represented by the second term. 
}
\label{profile}
\end{figure}

In order to discuss double formation, 
we consider a profile with $(A,B,\si_1,\si_2,n)=(0.99,0,1.45,10\si_1,1)$, 
depicted in Fig.\ref{profile}. 
In this case, the central perturbed region, represented by the first term of (\ref{newfunction}), 
is superposed on the perturbed region represented by the second term 
whose length scale is ten times larger than the central perturbed region. 
The first term itself corresponds to an initial perturbation which can collapse to form a PBH 
after the perturbed region $r\lesssim \sigma_1$ crosses the horizon as mentioned earlier. 
The perturbation represented by the second term is physically equivalent to the following profile, 
after a scale transformation $r\rightarrow \sqrt{1-A}r$:
\be
K_{\rm{i}}=\exp\left[-\left(\fr{r}{\sqrt{1-A}\sigma_2}\right)^{2}\right].
\ee
So when $\sqrt{1-A}\sigma_2=\si_1$, which holds in the current parameter choice, 
the perturbation represented by the second term 
is equivalent to the one represented by the first. 
Therefore, the second term itself can also collapse to form a PBH after horizon crossing without the presence of the first term. 
Physically, what is expected to happen from this initial set up is that 
the central region, represented by the first term, collapses to form a PBH as it enters the horizon 
and then the larger-scale perturbation represented by the second term collapses to form a larger PBH after this scale 
crosses the horizon, involving the central smaller PBH already formed earlier. 

\begin{figure}[h]
\begin{center}
\includegraphics[width=15cm,keepaspectratio,clip]{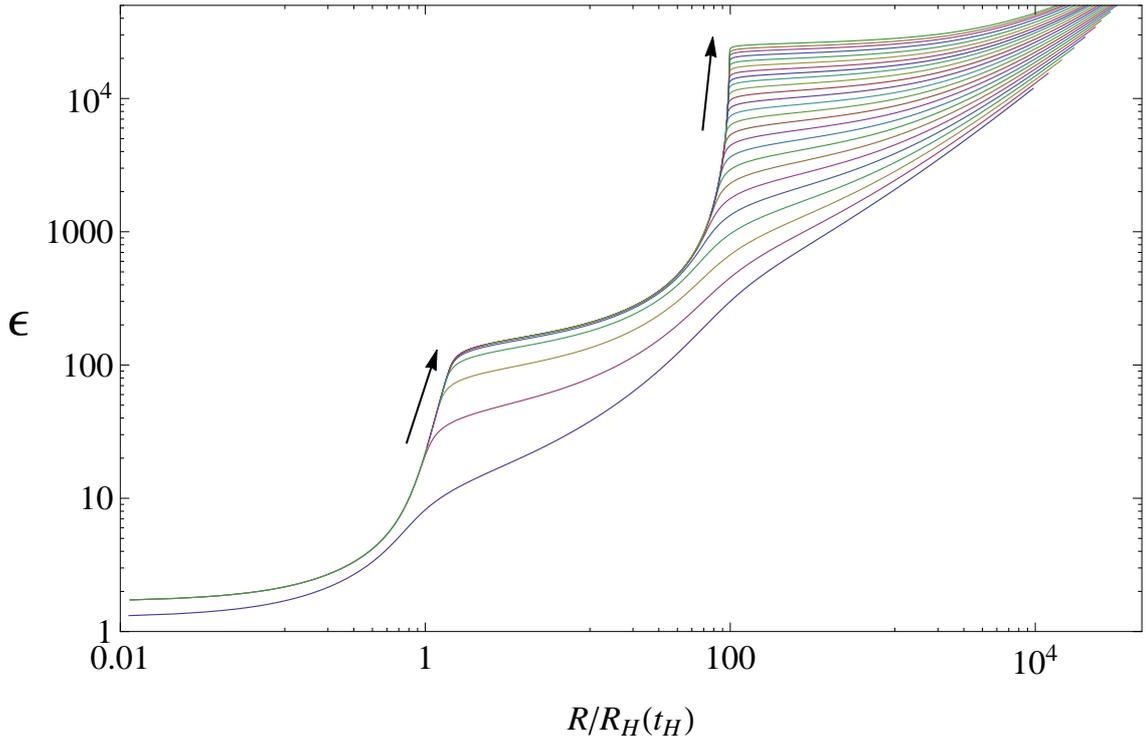}
\end{center}
\caption{Null geodesics of photons for the case of a double formation of PBHs 
where $(A,B,\si_1,\si_2,n)=(0.99,0,1.45,10\si_1,1)$. 
Arrows represent the direction of the time evolution. 
Photons emitted at later times first become 
almost trapped by the smaller PBH, and narrowly escape 
to the outer region, where they once more become almost trapped by the larger PBH 
before they escape to infinity. 
}
\label{double1}
\end{figure}

\begin{figure}[h]
\begin{center}
\includegraphics[width=15cm,keepaspectratio,clip]{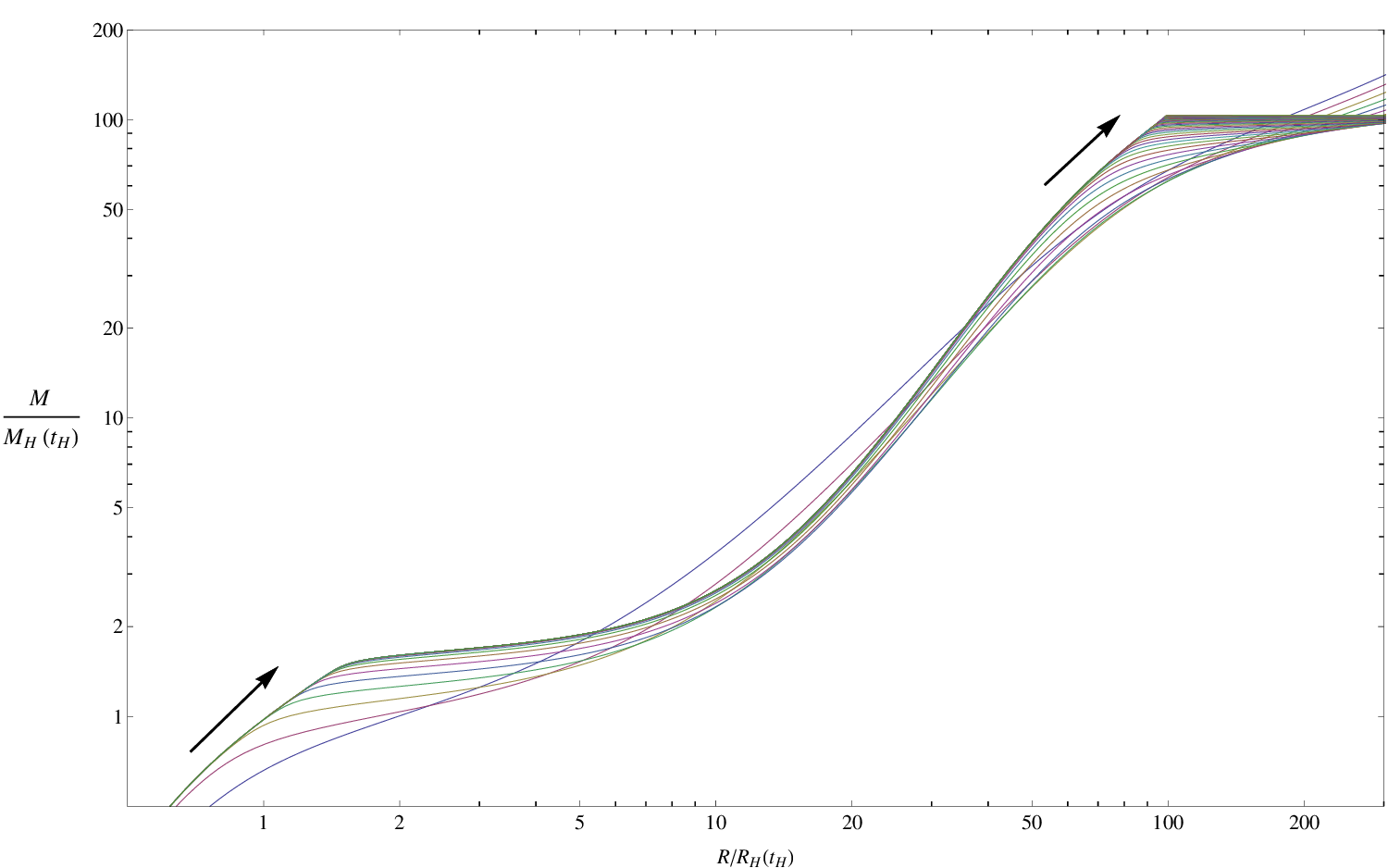}
\end{center}
\caption{The time evolution of the mass profile for the case of a double formation of PBHs 
where $(A,B,\si_1,\si_2,n)=(0.99,0,1.45,10\si_1,1)$. 
The arrows represent the direction of the time evolution. 
The two flat parts measure the mass of the smaller PBH and larger one, respectively}
\label{double2}
\end{figure}

We confirm this prediction by a numerical computation with the aforementioned initial curvature profile provided as the initial condition; 
results are shown in Fig.\ref{double1} and \ref{double2}. 
First, a PBH with mass around $1.5M_{\rm{H}}(t_{\rm{H}})$, where $t_{\rm{H}}$ is the horizon-crossing time defined by the first term of (\ref{newfunction}), is formed, and then 
another larger PBH $\sim 100M_{\rm{H}}(t_{\rm{H}})$ is formed. 
Note that the mass of the smaller PBH is somewhat larger than $M_{\rm{H}}(t_{\rm{H}})$, 
the mass of the PBH in the previous case with $(A,B,\si_1,n)=(1,0,1.45,1)$, 
even though the first term is equivalent to this case. 
This is due to the existence of the second term describing the larger scale perturbation, 
which makes the average density around the central region larger at the time of the formation of the smaller PBH. 
On the other hand, the mass of the larger PBH is almost 100 times larger than the previous case with $(A,B,\si_1,n)=(1,0,1.45,1)$, 
which can be understood as follows. 
First of all, in this simulation of double formation, 
the radius of the overdense region $r_{\rm{i}}$ is defined by the first term of (\ref{newfunction}). 
So let us denote this radius by $r_{\rm{i},1}$ 
to be contrasted with $r_{\rm{i},2}$, the radius of the overdense region 
defined by the second term. 
Since $r_{\rm{i},1}\propto \si_1$ and $r_{\rm{i},2}\propto \si_2$, 
we find $r_{\rm{i},2}=10r_{\rm{i},1}$. 
Then, denoting the horizon crossing time defined by the first term as 
$t(\ep(r_{\rm{i},1})=1)$, we have $t(\ep(r_{\rm{i},2})=1)=100t(\ep(r_{\rm{i},1})=1)$ from eq.(\ref{defofep}). 
Since the Hubble radius and the horizon mass are proportional to $t$, 
we find $M_{\rm{H}}(t(\ep(r_{\rm{i},2})=1))=100M_{\rm{H}}(t(\ep(r_{\rm{i},1})=1))$ as well as 
$R_{\rm{H}}(t(\ep(r_{\rm{i},2})=1))=100R_{\rm{H}}(t(\ep(r_{\rm{i},1})=1))$. 
Hence, the upper right parts of Fig.\ref{double1} and \ref{double2} are obtained by rescaling 
the left and right panel of Fig.\ref{single} by $\sim 100$ respectively, since 
the vertical and horizontal axis of the Fig.\ref{double1} and \ref{double2} are normalized by 
the horizon mass and the Hubble radius at the time of the 
crossing \textit{defined by the first term} of eq.(\ref{newfunction}).


\section{The effects of high-frequency modes}
In numerical simulations of the formation of PBHs, 
the presence of high-frequency modes (hereafter HF modes), whose wavelength is much shorter than the perturbed region under consideration, 
are not taken into account \cite{Shibata:1999zs,Polnarev:2006aa,1979STIN...8010983N,PhysRevD.59.124013,1475-7516-2014-01-037}. 
HF modes, however, should exist since in principle the power spectrum of primordial curvature perturbations has an extended profile 
and thus affect the formation of PBHs to some extent. 
In this section, the effects of HF modes are discussed. 

\begin{figure}[h]
\begin{center}
\includegraphics[width=10cm,keepaspectratio,clip]{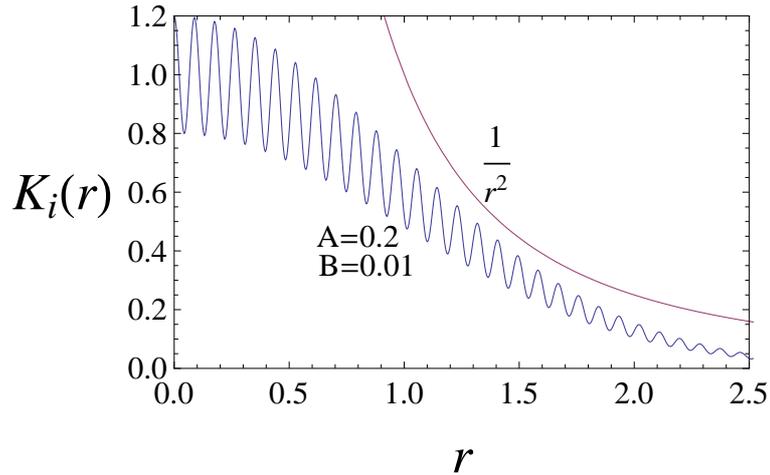}
\end{center}
\caption{An example of the initial curvature profile 
described by eq.(\ref{HF}).
}
\label{hfprofile}
\end{figure}

To this end, let us introduce the following initial curvature profile:
\be
K_\mathrm{i}(r)=
\exp\left[-\left(\fr{r}{\sigma_1}\right)^{2}\right]
\left[1+A\cos\left(\fr{r}{B\sigma_1}\right)\right].
\label{HF}
\ee
When $B<1$, this function represents situations where 
a HF mode is superposed upon a perturbation of longer wavelength, 
as is shown in Fig.\ref{hfprofile}. 

\begin{figure}[h]
\begin{center}
\includegraphics[width=15cm,keepaspectratio,clip]{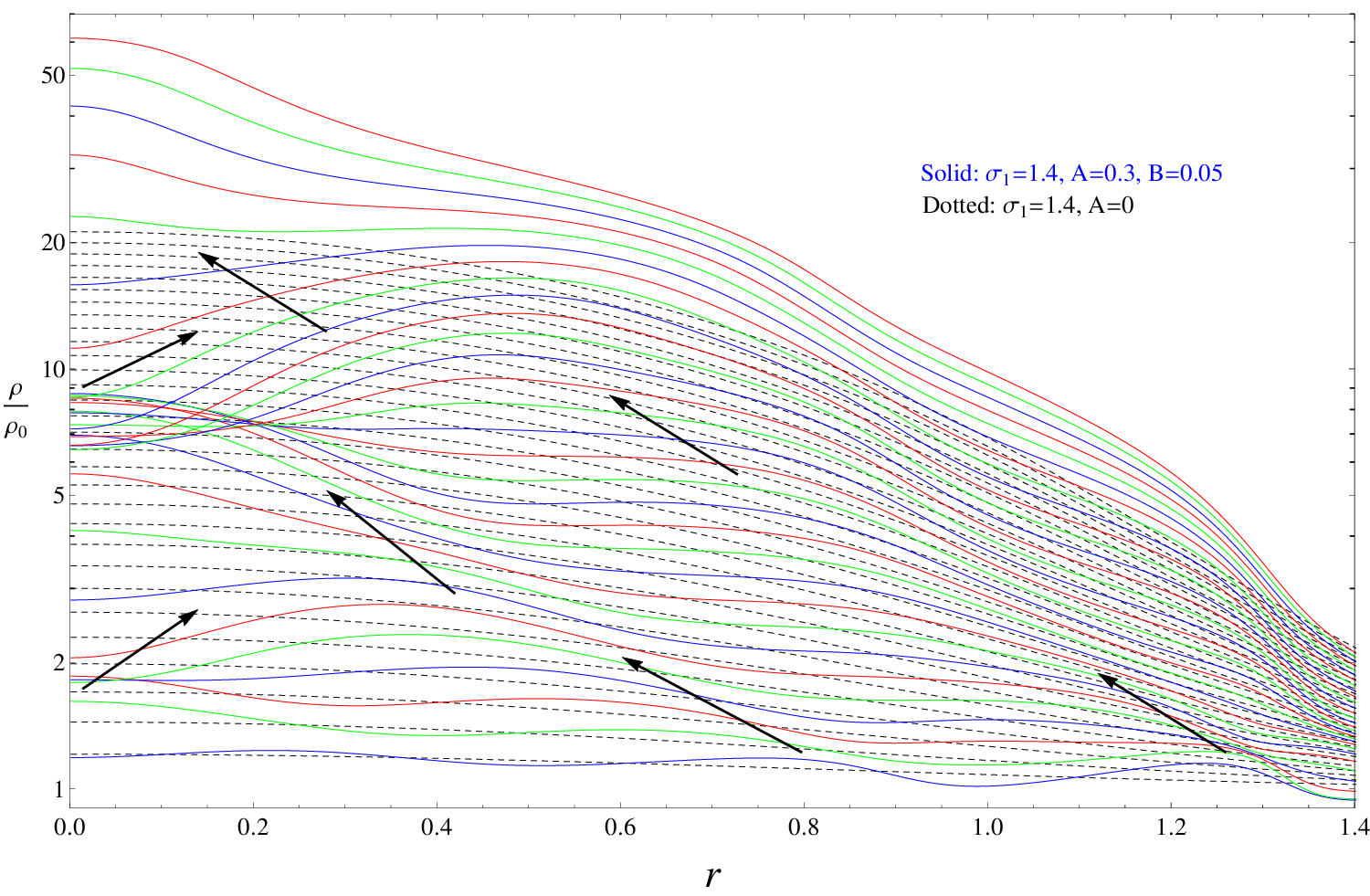}
\end{center}
\caption{An example of the time evolution of the density perturbation for a case where 
a PBH is eventually formed. 
Each curve corresponds to the density perturbation profile at $\epsilon=0.1,0.2,\cdots,2.9,3.$ 
For comparison, the density perturbation profile for the same time sequence for a case with the same value of $\si_1$ but without the HF mode, 
in which case a PBH can not be formed,  
is shown by the dashed lines. 
The arrows indicate the direction of the fluid motion. 
}
\label{evolution}
\end{figure}

The time evolution of the energy density perturbation of a typical case is shown in Fig.\ref{evolution}. 
The HF mode crosses the horizon first and starts to grow before the main, or long-wavelength perturbation crosses the horizon. 
At this point, the main perturbation does not seem present as long as we focus on the \textit{density} perturbation, 
since the density perturbation is suppressed on super-horizon scales in the comoving slicing we employ. 
After the horizon crossing of the main perturbation, 
the density perturbation with the corresponding wavelength starts to grow, and
the HF mode starts to propagate towards the center due to stronger gravity in the center, resulting from the main perturbation. 
When a local maximum arrives at the center, it bounces, but soon it pulls back towards center once more and as a whole the energy density in the center seems to increase more 
rapidly than the case without the HF mode. 
As a result, 
the value of the density perturbation at the center fluctuates significantly, as is shown in Fig.\ref{evolution}
\footnote
{
This makes it difficult to determine (as quickly as possible to reduce computational costs) 
when the perturbation is destined to die without forming 
a PBH, 
the determination which is necessary to 
investigate the formation condition of PBHs. 
Without the presence of HF modes, determining when a perturbation is destined to vanish is simple, 
since in this case the density perturbation at the center 
monotonically increases when a BH is eventually formed, and once it starts to decrease, 
the perturbation will definitely die so at this time one can stop numerical integration. 
In contrast, when a HF mode is present, one cannot conclude the perturbation will decay even if the density perturbation at the center 
starts to decrease, because it can be due to the presence of the HF mode, as is shown in Fig.\ref{evolution}. 
So careful analysis is required to ensure the quasi-global decrease in the density perturbation before stopping numerical integration in cases including HF modes. 
}.

\begin{figure}[h]
\begin{center}
\includegraphics[width=14cm,keepaspectratio,clip]{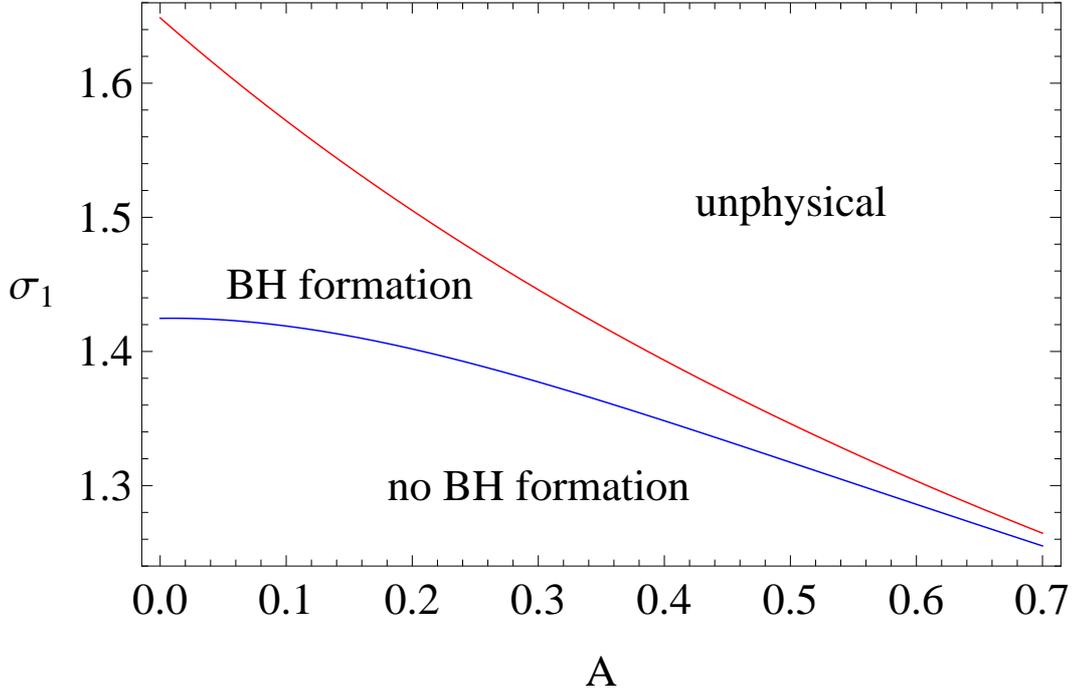}
\end{center}
\caption{The PBH formation condition for the initial curvature profiles 
described by eq.(\ref{HF}) with $B=0.03$. The "unphysical" region is determined by the condition eq.(\ref{condforKi})
}
\label{condition}
\end{figure}

In Fig.\ref{condition}, the PBH formation condition for this class of profiles is shown, where $B$ is fixed to $0.03$. 
For $A=0$ (without the HF mode), the PBH is formed when $1.42\lesssim\sigma_1$. 
When $A$ is larger, the threshold value decreases, implying the HF mode somewhat \textit{facilitates} the formation of PBHs, 
though one may have expected the HF mode to hinder formation. 
One may naively suspect such facilitation is simply because, 
first assuming the local maxima to help PBH formation and local minima 
to hinder formation, inner local extrema are more effective than outer local extrema 
and (\ref{HF}) implies the innermost local extremum is a local maximum if $A$ is positive, hence the effects of local maxima dominate, leading to a 
decrease of the threshold for larger values of $A$. 
It turns out that this is not the case by conducting numerical simulations for negative values of $A$ and recovering Fig.\ref{condition}, 
with the horizontal axis replaced by the absolute value of $A$. That is, the phase of the HF mode is not important. 

To understand why HF modes help the formation of PBHs, let us look at Fig.\ref{evolution} once more, 
showing the local maxima start to move towards the center after horizon crossing of the main perturbation. 
This behavior seems to result from strong gravity in the center due to the main perturbation. 
This indicates more effective transportation of radiation towards the center, which may explain 
the reason for the decrease in the threshold value when a HF mode is present. 

In Fig.\ref{condition}, the blue threshold line seems to converge to the line defining the "unphysical" region, determined by the condition eq.(\ref{condforKi}). 
However, what happens for even larger values of $A$ is difficult to investigate due to large spatial and time derivatives near the center. 

It also turned out that 
the threshold is insensitive to 
the wavelength of the HF mode (confirmed in the range $0.01<B<0.2$), and that 
introducing two HF modes at the same time facilitates PBH formation somewhat more. 

\section{Conclusion}
First, the double formation of PBHs is discussed, where 
a smaller PBH is swallowed by another bigger PBH. 
Suppose there exists a highly perturbed region which will collapse to form a PBH after the horizon crossing of this region, 
and also that this region is superposed on much 
larger region, which also collapses, as it enters the horizon later. 
Then, what should happen is the collapse of the central smaller region at the time of the crossing of this 
region, which is followed by another collapse of the larger perturbation at the time of the crossing of this 
larger perturbation. The smaller PBH, formed earlier, is swallowed in the second collapse leading to a single 
larger PBH as the final state. The first collapse turns out to be insensitive to the presence of the larger perturbation 
since the larger perturbation is still outside the horizon at that moment. 
In addition, the second collapse is not affected by the already formed small PBH due to the large scale difference. 
In this paper we have reported a first numerical confirmation of this phenomenon of the double PBH formation. 

Second, the effects of HF-modes on the formation of PBHs are discussed. 
This issue has been numerically investigated for the first time and, we find that 
HF-modes facilitate the formation of PBHs, decreasing the threshold value required for the formation of PBHs. 
This could potentially increase the abundance of PBHs by several orders of magnitude. 
This is because local small-scale overdensities, superposed on a larger perturbation, 
fall into the center due to the strong gravity realised by the larger perturbation, 
leading to more efficient transportation of radiation towards the center. 

These results show that the calculation of probability distribution of primordial inhomogeneities is essential to 
precisely predict the abundance of PBHs.

\section*{ACKNOWLEDGMENTS}
This work was partially
supported by Grant-in-Aid for JSPS Fellow No. 25.8199.
The author thanks Jun'ichi Yokoyama for useful comments, reading the manuscript, and continuous encouragement.
The author also thanks Kevin Croker for useful comments.

\bibliography{bib205}

\end{document}